\documentclass[twocolumn,preprintnumbers,floatfix,superscriptaddress,noshowpacs,aps,pra]{revtex4-2}

\usepackage[mathletters]{ucs}
\usepackage[utf8x]{inputenc}
\usepackage{microtype}

\usepackage{amsmath}
\usepackage{amssymb}
\usepackage{bbm}

\usepackage[hidelinks]{hyperref}
\usepackage{multirow}
\usepackage[sort&compress]{natbib}

\usepackage{graphicx}
\usepackage[export]{adjustbox}
\usepackage{hhline}

\usepackage{float}
\usepackage{placeins}
\usepackage{booktabs}
\usepackage{subcaption}
\usepackage{makecell}
\usepackage{tabstackengine}

\usepackage{xspace}
\usepackage[dvipsnames]{xcolor}

\usepackage{xspace}
\usepackage{xfrac}
\usepackage{hyperref}
\usepackage[nameinlink]{cleveref}
\usepackage{appendix}
\usepackage{units}
\usepackage{xifthen}
\usepackage{xcolor}
\hypersetup{
	colorlinks,
	linkcolor={blue!75!black},
	citecolor={blue!75!black},
	urlcolor={blue!75!black}
}

\usepackage[normalem]{ulem}

\newcommand{\sgn}{\mbox{sgn}}
\newcommand{\Tr}{\mathrm{Tr}}
\def\CV{{\mathcal V}}

\newcommand{\imag}{\text{i}}

\hypersetup{
  colorlinks,
  linkcolor={red!75!black},
  citecolor={blue!75!black},
  urlcolor={blue!75!black},
  pdftitle={Towards quantitative precision in ultracold atoms with functional renormalisation},
  pdfauthor={Faigle-Cedzich, Pawlowski, Wetterich},
  pdfkeywords={RG} {Equation of state} {Gap} {BCS-BEC crossover},
  bookmarksopen=true,
  bookmarksopenlevel=2,
  bookmarksnumbered=true
}

\graphicspath{{./figures/}}
\begin{document}

\title{Towards quantitative precision in ultracold atoms with functional renormalisation}
%\date{\today}
\author{Bruno M. Faigle-Cedzich}
\affiliation{Institute for Theoretical Physics, Heidelberg University, D-69120 Heidelberg, Germany}
\author{Jan M. Pawlowski}
\affiliation{Institute for Theoretical Physics, Heidelberg University, D-69120 Heidelberg, Germany}
\affiliation{ExtreMe Matter Institute EMMI, GSI Helmholtzzentrum für Schwerionenforschung mbH, D-64291 Darmstadt, Germany}

\author{Christof Wetterich}
\affiliation{Institute for Theoretical Physics, Heidelberg University, D-69120 Heidelberg, Germany}

\begin{abstract}
We compute the equation of state, the gap as well as the density fluctuations of a two-component superfluid Fermi gas over the whole range of BEC-BCS crossover at vanishing temperature within the functional renormalisation group approach. With an improved understanding of the relation between density and chemical potential, already a rather simple truncation yields a very good quantitative agreement with experimental data and theoretical results, in particular in the unitarity limit and on the BEC side. The current approach utilises higher order density fluctuations as a  fundamental building block for the computation of the density as a function of the chemical potential. This circumvents the fine-tuning problem of the density-related fundamental parameters on the microscopic level that has been observed in previous approaches. The quantitative reliability of the functional renormalisation group approach already within simple approximations opens the path towards precision results in more elaborate truncations. 
\end{abstract}

%\pacs{}

\maketitle

%\tableofcontents

%%%%%%%%%%%%%%%%%%% Content %%%%%%%%%%%%%%%%%%%%

\section{Introduction}

The phase structure of ultracold Fermi gases, in particular the resolution of the equation of state and change of observables such as the gap through the BEC-BCS crossover, has been studied intensively in the past decades. Ultracold Fermi gases show interesting macroscopic quantum phenomena such as superfluidity and, thanks to there amazing tunability, are also tailor made as model systems for many non-relativistic and relativistic systems. Precision results for these systems constitutes a challenge for the development of theoretical approaches beyond perturbation theory. 

The system has been investigated with the functional renormalisation group (fRG) approach in the past two decades, and many of its interesting phenomena have been explained within this approach, for respective reviews see e.g.~\cite{Kopietz:2010zz, Metzner:2011cw, Scherer:2010sv, Boettcher:2012cm, Dupuis:2020fhh}. In the context of the BEC-BCS cross-over, many observables agree well with experiment as well as other theoretical approaches. In particular, the fRG approach has a very good grip on the physics phenomena at work, covering all relevant limits within a global approach. On the other hand, the quantitative precision of the equation of state (EoS), encoded in the chemical potential dependence of the density, remained poor despite the elaborate truncation level achieved in the past two decades.  This affects observables often used for benchmarking theoretical methods as the Bertsch parameter or the superfluid gap. At first sight this shortcoming may seem surprising, also given that these discrepancies prevail in elaborate approximation schemes.  We will see that this issue originates from a fine tuning problem for density-related microscopic parameters, which can be circumvented by a more careful setting.   

In the present work we show that the fRG-approach to ultracold gases indeed offers already quantitative precision even in relatively simple approximations, that capture the dynamics of the systems very well. The discrepancies  for the above mentioned observables such as the Bertsch parameter is solely related to an insufficient resolution of specific density observables with a positive momentum dimension. As none of the density-related correlation functions feeds back into the flow of the couplings at fixed chemical potential $\mu$, this does not affect the quantitative precision obtained for the effective action and its derived observables. 

For the resolution of the density observables including the Bertsch parameter we put forward an fRG approach, in which these observables are obtained from an interactive $\mu$-integration of density fluctuations and hyper fluctuations, which are the $n$th derivatives of the density $n(\mu)$ w.r.t.~$\mu$. These derivatives successively lower the ultraviolet sensitivity of the fluctuation observables. In short, for a sufficiently large order, the hyper fluctuations show the required $\mu$-independence for large momentum scales $k$ and decay with increasing momentum cutoff scale $k$. Accordingly, their momentum cutoff fRG flow can be integrated without an initial value or fine-tuning problem, which would be present for a direct solution of the flow equation of $n(\mu)$. This scheme has been set-up in the context of low energy QCD in \cite{Fu:2015naa}, and a first application to ultracold atoms can be found in \cite{Faigle-Cedzich:2019uvr}. For ultracold gases it suffices to start with the skewness $\partial_\mu^2 n$, the third $\mu$ derivative of the free energy of the system. Then,  density-related observables such as the Bertsch parameter are obtained via $\mu$-integrations and yield quantitative results within a systematic approach.  

For the explicit numerical computation we use a relatively simple approximation of the full two-component Fermi gas, that nonetheless already provides quantitative access to the the BEC-side and the unitary limit. Its improvement on the BCS side has been studied at length in the literature, see~\cite{Kopietz:2010zz, Metzner:2011cw, Scherer:2010sv, Boettcher:2012cm, Dupuis:2020fhh}. The present approximation is easily extended in this direction, but this is beyond the scope of the present work. The present quantitative results within this relatively simple approximation emphasise that the fRG approach is well-suited for the access to ultracold atomic systems as it captures the respective physics in a uniform manner in the whole crossover regime.

%%%%%%%%%%%% Model & fRG %%%%%%%%%%%%%%%%%
\section{Ultracold Fermi gases with the functional renormalisation group}
\label{sec:model}

In this Section we briefly introduce the fRG set-up used here. More details and results can be found in~\cite{Faigle-Cedzich:2019uvr} and the reviews~\cite{Kopietz:2010zz, Metzner:2011cw, Scherer:2010sv, Boettcher:2012cm, Dupuis:2020fhh} and literature therein. 
 
The functional renormalisation group approach approach is based on the flow equation of the effective action (Gibbs free energy) of the system. The microscopic action of ultracold fermionic gases such as ${}^6\text{Li}$ and ${}^{40}\text{K}$ is well described by a pointlike scattering interaction between the fermionic atoms and their dispersion term, 
\begin{align}\nonumber 
	S[\psi]=&\int_X\Biggl\{\sum_{\sigma=1,2}\psi_\sigma\,(\partial_\tau-\nabla^2-{\mu})\,\psi_\sigma\\[1ex]
&\hspace{2cm}	+\lambda_{\psi} \psi^*_1 \psi^*_2 \psi_2 \psi_1 \Biggr\}\,,
\label{eq:action_micro}
\end{align}
with $\hbar=k_B=2M_\psi=1$. The fermions $\psi^T=(\psi_1,\psi_2)$ correspond to two hyperfine states and  
\begin{align}
	\int_X=\int_0^\beta d\tau \int d^{d}x\,,\qquad X=(\tau,\vec{x})\,. 
\end{align}
The four-Fermi coupling is related to the scattering length, $\lambda_\psi=8\,\pi\,a$. The temperature $T$ enters by periodicity of the $\tau$-integral with $\beta=1/T$, and we consider the zero temperature limit $\beta\to \infty$. 

The quantum analogue of the microscopic action \labelcref{eq:action_micro} is the quantum effective action $\Gamma[\psi]$, that depends on the macroscopic (mean) field $\psi$. Ultracold atomic gases experience a phase transition from the atomic phase to a condensate phase at sufficiently low temperature. The condensate phase includes a BCS-type regime and a BEC regime that are separated by the Feshbach resonance at diverging scattering length. 

On the BEC side the dynamics of the theory is that of tightly bound molecules (dimers) $\phi = \psi_1\psi_2$. Its is advantageous to incorporate these low temperature degrees of freedom as effective fields via a Hubbard-Stratonovich transformation 
\begin{align}
	S=\int_X &\left[\sum_{\sigma=1,2} \psi^{*}\left(\partial_\tau-\nabla^2-\mu\right)\psi\right.\nonumber\\[1ex]
	&+m_\phi^2\,\phi^*\phi-\left.h\,\left(\phi^*\,\psi_1\,\psi_2-\phi\,\psi_1^*\,\psi_2^*\right)\vphantom{\sum_{\sigma=1,2}}\right]\,. 
\label{eq:action_HS}
\end{align}
Then, the functional integral also includes an integration over the dimer field $\phi$. Upon a Gau\ss ian integration over $\phi$, the action \labelcref{eq:action_HS} is equivalent to the microscopic action \labelcref{eq:action_micro} with 
\begin{align}
	\lambda_\psi=-\frac{h^2}{m_\phi^2}\,.
\end{align}
The Feshbach coupling $h$ accounts for the scattering of two fermionic atoms $\psi$ with different spin into a bosonic dimer $\phi$.

\subsection{Flow equation of the effective action}

The microscopic action in \labelcref{eq:action_micro} or its bosonised version \labelcref{eq:action_HS} governs the atomic physics at microscopic momentum scales $k\propto \Lambda$, where $\Lambda^{-1}$ is of the order of the Van der Waals length $\ell_{\text{VdW}}$. The system enters the superfluid phase at much smaller momentum scales, $k\ll \Lambda$. The respective thermal and quantum fluctuations can be incorporated with the functional renormalisation group. In the fRG approach an infrared momentum cutoff scale $k$ is introduced via a modification of the fermionic and bosonic dispersions. Below  the cutoff scale quantum and thermal fluctuation are suppressed, for details see \Cref{app:TechnicalDetails}. The infinitesimal change of the effective action $\Gamma_k$ with respect to the momentum scale $k$ is governed by the flow equation \cite{Wetterich:1992yh, Diehl:2007ri}, 
\begin{align}\label{eq:flowG}
	\partial_t \Gamma_k[\psi,\phi]= \frac12 \Tr \,G_{k,\phi} \,\partial_t R_{k,\phi}
	- \Tr \,G_{k,\psi}\,\partial_t R_{k,\psi}\,,
\end{align}
with $\partial_t = k\partial_k$. The infrared regularised propagators $G_k[\psi,\phi]$, 
\begin{align} 
	G_k[\psi,\phi] = \frac{1}{\Gamma^{(2)}_k[\psi,\phi]+R_k}\,,
\end{align}
involve $\Gamma^{(2)}_k$, the second derivative where $\Gamma_k$ w.r.t.\ to $\psi$ and $\phi$. 

In the present work we consider a very simple Ansatz for the effective action in \labelcref{eq:flowG}, which still  suffices to provide us with quantitative results for the gap and the equation of state (EoS), 
\begin{align}\nonumber 
	\Gamma_k[\psi,\phi] = &\, \\[1ex]\nonumber 
&\,\hspace{-1cm}	\int_{X} \Biggl\{   \psi^*\Bigl[ \left( \imag\,q_0 -\mu\right) +q^2\Bigr] \psi+ \phi^*\,\Biggl[  S_{\phi}\,\imag\,q_0+\frac{q^2}{2} \Biggr] \phi\\[1ex]
	& +U\left(\rho\right)-h\,\Bigl(\phi^*\,\psi_1\,\psi_2 - \phi\,\psi_1^*\,\psi_2^*\Bigr)\Biggr\}\,, 
\label{eq:Gk}
\end{align}
with $\rho = \phi^*\phi$. The second line combines the kinetic terms of fermionic atoms and bosonic dimers. We have dropped the wave function of the fermions and also consider the Yukawa coupling $h$ as $k$-independent. We concentrate on the non-trivial low temperature dynamics of the dimer, which is encoded in the effective potential $U(\rho)=U_{k=0}(\rho)$. In particular, the superfluid phase is entered for a non-trivial minimum $\rho_0$ of $U(\rho)$. Apart from the flow of $U_k(\rho)$, we only follow the $k$-dependence of the parameter $S_\phi$.   

The flow of these quantities is supplemented by the anomalous dimension that originates from the spatial wave function $A_\phi$. It is convenient to absorb the spatial wave function $A_{\phi}$ into the bosonic field, 
\begin{align}
 A_\phi \phi \to \phi\,. 
\end{align}
Then, $S_{\phi}=Z_\phi/A_\phi$ is a ratio of the temporal wave function $Z_\phi$ and spatial wave function $A_\phi$, while the spatial wave function has dropped out, its impact is present in the flow via its anomalous dimension. Note that all other coupling parameters are also rescaled with respective powers of $A_\phi$. In particular we have $h\to h/A^{1/2}_\phi$ and the latter is kept fixed in the current approximation. This simulates effectively the common behaviour of Yukawa couplings in flows with dynamical condensation put forward in  \cite{Gies:2001nw}, see also \cite{Pawlowski:2005xe, Floerchinger:2009uf, Fu:2019hdw}. 

The chemical potential enters \labelcref{eq:Gk} as  imaginary shift of the frequency $q_0$. The chemical potential in the bosonic term has been absorbed in the mass term in the effective potential.For the latter we consider the-$\phi^4$ approximation, 
\begin{align}
U(\rho) =  m_\phi^2   (\rho-\rho_{0,k} ) +\frac{\lambda_\phi}{2} (\rho-\rho_{0,k} )^2\,,
\end{align}
and neglect the higher order terms $\rho^n$ with $n>2$. This approximation is well justified in regimes where dimer re-scatterings are of subleading nature, which holds true for most of the phase diagram. The explicit flows for the effective potential and the wave functions are deferred to \Cref{app:TechnicalDetails}. For more details as well as systematic extensions of the approximation used here see e.g.~\cite{Faigle-Cedzich:2019uvr, Boettcher:2012cm}.

%%%%%%%%%%%%%%%%%%%%%%%%%%%%%%%%%
\subsection{Dimensions and relevant parameters}
\label{sec:Dim+Parameters}

In the functional integral approach the system of ultracold atoms is described as a non-relativistic quantum field theory. As such, it contains a few \textit{relevant parameters}, which have to be fixed by observables. If the fRG flow is initiated at sufficiently large scales $k=\Lambda$, all other observables can be computed in terms of these few relevant parameters. The latter are related to the couplings of the microscopic theory for $k\to \Lambda$. In turn, their final macroscopic values for $k\to 0$ are dominated by the flow at large $k$ that contains positive powers and logarithms of $k$. Hence, finding the correct microscopic values of the relevant parameters typically involves the solution of a fine-tuning problem.  

Such a fine-tuning problem is present for the relation $n(\mu)$, the chemical potential dependence of the density. The core of this problem can already be seen by a simple dimensional analysis, where we count the dimension in powers of the (momentum) cutoff scale $k$. For non-relativistic quantum field theories with $2M_\psi=1$, the frequency $q_0$ scales as spatial momentum squared, $\vec q^2$.  This implies that $q_0$ and $\mu$ have the momentum dimension two and hence scale with $k^2$. The space-time integral has the momentum scaling $k^{-5}$, while the fermionic and bosonic fields scale with $k^{3/2}$. In consequence, the effective potential $U$ scales with $k^5$, which is the scaling of the field-independent constant part $U_0=U(\rho_0)$, where $\rho_0$ is the solution of the equation of motion. The pressure of the theory is given by $U_0$ which is a highly relevant parameter. The $\rho^n$ couplings in $U$ have a successively lower relevance, already the linear term in $\rho$ scales with $k^2$, as does its flow $\partial_t ( \partial U/\partial\rho)\propto k^2$. This $\rho^1$ coupling is simply a mass parameter $m^2_\phi= \partial U/\partial\rho$. The respective physical parameter or observable is the scattering length $a$  with 
\begin{align}
	a= -\frac{h^2}{8\pi\,\nu(B)}\,,
\end{align}
where the physical detuning is given by ${\nu(B) = \nu_{\Lambda}-h^2/(6\pi^2)\,\Lambda}$ and $h=h_{\Lambda}$.\\
Already the next $\rho$-derivative scales with $k^{-1}$, and is an irrelevant parameter. Its flow is dominated by the infrared contributions and its macroscopic value is determined by the relevant parameters. 

This leaves us with two relevant parameters in the effective potential. While $m_\phi^2$ enters the flow equation of the other couplings, the value of the relevant parameter $U_0$ decouples from that of the other couplings: the right hand side of the flow equation of all couplings depends on $\Gamma^{(2)}_k$ and higher correlation functions $\Gamma^{(n)}_k$. The zero- and one-point functions $\Gamma_k^{(0)}=\Gamma_k=\int_X U_0$  and $\Gamma^{(1)}_k$ drop out. However, $U_0$ matters for the density, which is given by 
\begin{align}
	n(\mu) =\frac{\partial U(\rho_0)}{\partial\mu }\,.
\label{eq:DensityDim} 
\end{align}
In the superfluid regime, $\rho_0$ is the solution of the equation of motion, $U'(\rho_0)=0$, with $U'=\partial U/\partial\rho$. In turn, in the symmetric regime with $m_\phi^2>0$ one has $\rho_0=0$. The above power counting entails that the flow of the density scales with $k^3$ in the UV, and hence the density is a relevant parameter. Accordingly, it needs to be fine-tuned with cubic precision in order to achieve a vanishing density for $\mu=0$. This cubic fine-tuning problem implies, that even the next derivative, the density-density correlation $\partial n/\partial \mu$, is relevant: it scales linearly with $k$ as $\partial^2 U_0/{\partial\mu^2}\propto k$. The situation changes only for the third derivative $\partial^3 U_0/ {\partial\mu^3} \propto k^{-1}$, which is $\partial^2 n/{\partial\mu^2}$. It has the desired UV-irrelevance and is dominated by its infrared flow. It can therefore be computed reliably without any UV--fine-tuning problem. One may safely set 
\begin{align}
	\left.\partial^3 U_0/ {\partial\mu^3} \right|_{k=\Lambda}\approx 0\,, 
\end{align}
at the initial cutoff scale $\Lambda$. Having reliably computed $\partial^2 n/{\partial\mu^2}$ for $k=0$, one can find $\partial n/{\partial\mu}$ and the density $n(\mu)$ by successive $\mu$-integrations of this quantity. The integration constants fix the corresponding relevant parameters without the need to follow their flow explicitly. No fine-tuning is needed any more. 

We conclude that the fine-tuning problem for relevant parameters is inherent to any investigation of the direct flow of the density. Furthermore, this fine-tuning depends on the truncation. This problem for microscopic parameters related to the density is present even though the density is physically dominated by occupation numbers for low momenta close to the Fermi momentum. Our proposal avoids fine tuning problems and therefore opens the path for a precise determination of $n(\mu)$, which was missing so far in the fRG approach.

%%%%%%%%%%%%%%%%%%%%%%%%%%%%%%%%
\subsection{Hyper-fluctuations and density observables in the fRG approach}
\label{sec:density}

The density is given by the chemical potential derivative of the effective action at vanishing cutoff scale $k=0$, evaluated on the equations of motion for the fields, 
\begin{align}
	n= \frac{1}{\CV}\frac{d\,\Gamma[0,\phi_\textrm{\tiny{EoM}}]}{d\mu}\,, 
\label{eq:density} 
\end{align}
with the spatial volume $\CV$ and $\psi_\textrm{\tiny{EoM}}=0$. For states with constant $\phi$ this reduces to \labelcref{eq:DensityDim}. The $\mu$-derivative hits both the explicit chemical potential dependence of the effective action and the implicit one in the couplings, as well as $\phi_\textrm{EoM}$. While the latter term drops out in \labelcref{eq:density} as it is proportional to the EoM, these terms contribute for higher order derivatives in $\mu$. 

One may try to obtain the density with the $\mu$-derivative of the effective action at $k=0$, and hence directly from \labelcref{eq:density}. This approach was followed in the past, which implies a vanishing  microscopic value of the density, $n_{k=\Lambda}=0$. This typically differs from the value required by the fine tuning. An additional shortcoming of this approach is that for simple truncations the values of the couplings at vanishing momentum are effectively used, while the density is dominated by modes with Fermi momentum. The integrated flow of $\partial_t n$ improves on this shortcoming, 
\begin{align}
	n= \int_\Lambda^0 \frac{d k}{k}  \partial_t n_k +n_\Lambda\,,\qquad \partial_t n = \frac{1}{\CV}\frac{d\,  \partial_t  \Gamma_k}{d\mu}\,. 
\label{eq:densityInt} 
\end{align}
\Cref{eq:densityInt} is the sum of the integrated flow and the initial density \labelcref{eq:densityInt}. Naïvely, the latter is the density at large spatial momentum scales, and vanishes accordingly. However, this heuristic interpretation relies on a direct interpretation of the cutoff scale $k$ with a physical momentum scale. Setting $n_\Lambda=0$ leaves us with  the integrated microscopic contributions to the flow $\partial_t n_{k\to \Lambda} \propto \Lambda^3$. This part of the integrated flow is cancelled by the initial condition $ n_\Lambda\propto \Lambda^3$, otherwise we are left with a $\Lambda$-dependent result at $k=0$. This cancellation is called RG-consistency, see \cite{Braun:2018svj}.

For a quantitative understanding of the fine-tuning problem that can lead to discrepancies for density-related observables, we have a closer look at the density flow \labelcref{eq:densityInt} at asymptotically large cutoff scales $k$. The discussion of the general UV-scaling is deferred to \Cref{app:TechnicalDetails}: importantly, in  \Cref{app:UV-Scaling} it has been shown, that the common Fermi-surface regulators such as \labelcref{eq:flat_reg} amplify the UV fine-tuning problem by two orders. The following UV-scaling analysis is based on the UV-improved regulators discussed in \Cref{app:TechnicalDetails}. The specific UV-improved regulator used for the numerical results of the present work is provided in \labelcref{eq:CombinedRegulator} in \Cref{app:CombinedRegulator}.  

With the UV-improved regulator, the flow can be expanded in powers of the cutoff scale,
\begin{align}
  \lim_{k\to \infty} \partial_t n = c_{3}\, k^{3} + c_{1}\, \mu\, k
  + O\left(\frac1k\right)\,,
\label{eq:flown}
\end{align}
where $k\to\infty$ should be understood as the microscopic limit. A similar equation holds for the relativistic case, see \cite{Fu:2015naa}, where the limit entails $k$ being far larger than any mass scale in the problem. In the present case this leaves us with a cubic fine-tuning problem. Most importantly, the cubic fine-tuning problem includes a subleading linear one, being also  proportional to $\mu$. Fine-tuning problems of this type are difficult to resolve directly and we do this by considering the flows of susceptibilities. In particular, the flow of the second order susceptibility or variance $\partial_\mu n$ can be read of from \labelcref{eq:flown} in the large cutoff limit, 
\begin{align}
\lim_{k\to\infty}   \partial_t \frac{d \,n}{d \mu} = c_{1} k +O\left(\frac1k\right)\,,  
\label{eq:flowsus}
\end{align}
where we have used that total $t$- and $\mu$-derivative commute. Importantly, the linear fine-tuning problem in \labelcref{eq:flowsus} is independent of the chemical potential, only the terms decaying with at least $1/k$ may also depend on $\mu$. Accordingly, a further $\mu$-derivative eliminates all $\mu$-dependent terms that grow with the cutoff scale $k$. This leaves us with a flow for the third order fluctuation or skewness, that decays for large cutoff scales,
\begin{align}
 \lim_{k\to\infty} \partial_t \frac{d^2 \,n}{d \mu^2}  = O\left(\frac1k\right)\,. 
\label{eq:flowmusus}
\end{align}
In conclusion, the density $n$, and the second order susceptibility $d  n/d\mu $ at vanishing cut-off, $k=0$ can be obtained from integrating the third order susceptibility, which has a trivial initial condition for large cutoff scales. 
This can be done successively, and for the density we arrive at 
\begin{subequations}
	\label{eq:Intn}
\begin{align}
   n(\mu) = \int_0^\mu d\mu_1 \,  \frac{d \, n(\mu_1)}{ d\mu_1}\,,
  \quad \text{with} \quad 
   n(0)=0\,,
\label{eq:nbyintegral}
\end{align}
where we have used that the density vanishes at vanishing chemical potential. \Cref{eq:nbyintegral}  depends on the susceptibility $\partial_\mu n$, which has the integral representation 
\begin{align}\label{eq:susbyintegral}
  \frac{d  \,n(\mu)}{ d\mu} = \int_0^{\mu} d \,\mu_1\,  \frac{d^2  \,n(\mu_1)}{ d\mu_1^2}\,,
  \qquad \frac{d \, n(\mu)}{ d\mu}(0)=0\,.
\end{align}
\end{subequations}
It is left to determine $\partial_{\mu}^2\,n_k(\mu)$. Here we follow the approach put forward in \cite{Fu:2015naa}, where closed equations for the flows of $\mu$-derivatives of observables and couplings $\{g_i\}$ have been derived in a relativistic setting. This has been transferred to the present non-relativistic ultracold system in \cite{Faigle-Cedzich:2019uvr}. We recall the respective derivations in \Cref{app:density}, and summarise the results here: 

The set of all $k$-dependent parameters of the effective action are summarised in $\boldsymbol{g}=(g_1,...,g_N)$,  where the 'couplings' $g_i$ include the wave functions, masses, interaction strengths such as $\lambda_\phi,S_\phi,A_\phi$ and $m^2_\phi$ (symmetric phase) or $\rho_0$ (broken phase) in the present case. Note that in general the $g_i$ are fully frequency- and  momentum-dependent coefficients of the $\Gamma^{(n)}_k$. As indicated above, $\boldsymbol{g}$ also includes a $\mu$-dependent evaluation point $\phi_k(\mu)$ in the present expansion scheme. The latter 'coupling' replaces $m_\phi^2$ in the broken phase, where the flow is evaluated on the $k$- and $\mu$-dependent minimum. These closed equations allow to compute all flows in an iterative hierarchy from the flow of the effective action: the flow of $m$th order derivatives of observables can be computed from the knowledge of all $m-1$st order derivatives, see  \Cref{app:density}. We quote the respective result \labelcref{eq:flowni_App}, 
\begin{subequations}
\label{eq:flowni}
\begin{align}
		\frac{d  \,\dot{n}^{(i)} }{d \mu}=
		\left( \partial_\mu+
		\sum_{j=1}^{i+1} g_{m}^{(j)}
		\frac{\partial}{\partial_{g^{(j-1)}_{m}}}
		\right)\, \dot n^{(i-1)}\,,
		\label{eq:flowmoments}
\end{align}
with 
\begin{align}
\dot n^{(i)}=\dot n^{(i)}(\mu; \boldsymbol{g},..., \boldsymbol{g}^{(i+1)})\,.
\end{align}
In \labelcref{eq:flowmoments} we have used the notation $\dot f=\partial_t f$ and 
\begin{align}
	g^{(i)}_j=\frac{d  \,g^{(i-1)}_j}{d\mu}\,,\qquad \dot n^{(m)} = \frac{d \, \dot n^{(m-1)}  }{d \mu}\,,
\label{eq:mug}
\end{align}
and the lowest order  
\begin{align}
	g^{(0)}_j=g_j\,,\qquad \dot{n}^{(-1)}(\mu) = \dot\Gamma_k[0,\phi_\textrm{\tiny{EoM}}]\,.
	\label{eq:Lowest}
\end{align}
\end{subequations}
In conclusion, this leaves us with a closed set of differential equations for the density. With \labelcref{eq:flowmusus} it is obtained from integrating the third order susceptibility $n^{(2)}$ from $\mu=0$ with the initial conditions $n(0)=0$ and $n^{(1)}(0)=0$ to $\mu$. This requires the knowledge of $\boldsymbol{g}^{(i)}$ with $i\leq 3$, which can be iteratively computed from their flows, see	\labelcref{eq:Flowgn} in \Cref{app:density}. Both, $n^{(i)}$ and $\boldsymbol{g}^{(i)}$ are \textit{derived} quantities, that can be computed from the flow of effective action and do not feed back into it.

%%%%%%%%%%%%%%%%%%%%%%%%%%%%%%%%%%%
\section{Results}

We present results for the equation of state, \Cref{sec:EoS} and the gap, \Cref{sec:Gap} at vanishing temperature in comparison to theoretical and experimental results in the literature. A summary of these vacuum results for the EoS and the gap at unitarity are summarised in \Cref{tab:BertschGap}.

\begin{center}
	\begin{table}[b]
		\begin{tabular}{ |p{4.5cm}||p{1.75cm}|p{1.75cm}|  }
			\hline
			\multicolumn{3}{|c|}{Comparison of Bertsch parameter and gap at unitarity} \\
			\hline
			& $\mu/\epsilon_F$ & $\Delta/\epsilon_F$\\
			\hline
			fRG - this work & 0.362 & 0.392 \\
			fRG - effective potential & 0.55 & 0.60 \\
			\hhline {|=||=|=|}
			Literature results & & \\
			\hline
			Astrakharchik et al. \cite{Astrakharchik_2004} & 0.42(1) &  \\
			Hausmann et al.  \cite{Hausmann_2007} & 0.36 & 0.46 \\
			Bartosch et al. \cite{Bartosch:2009zr} & 0.32 & 0.61 \\
			\hline
			Experimental data &  & \\
			\hline
			Ku et al. \cite{Ku_2012} & 0.376(5) &  \\
			Schirotzek et al. \cite{Schirotzek_2008} &  & 0.44 \\
			Hoinka et al. \cite{hoinka_2017} &  & 0.47(3) \\
			Biss et al. \cite{biss_2021} & & \!\!$~$0.47(1)\\
			\hline
		\end{tabular}
		\caption{Comparison of the Bertsch parameter $\xi=\mu/\epsilon_F$ and the gap $\Delta/\epsilon_F$ at unitarity, $a\to \infty$, for vanishing temperature.}
		\label{tab:BertschGap}
	\end{table}
\end{center}

The fRG approach set-up here resolves the large discrepancy in the density-related observables observed so far. This shows that the fRG results in the literature have semi-quantitative reliability already in the simple approximation used here.  Earlier results for more elaborated truncations can partly be taken over. The only modification concerns the discrepancy due to the inaccurate determination of density-related observables done so far. 
\begin{figure*}[t]
	\centering
	\begin{subfigure}{.48\linewidth}
		\centering
		\includegraphics[width=.95\textwidth]{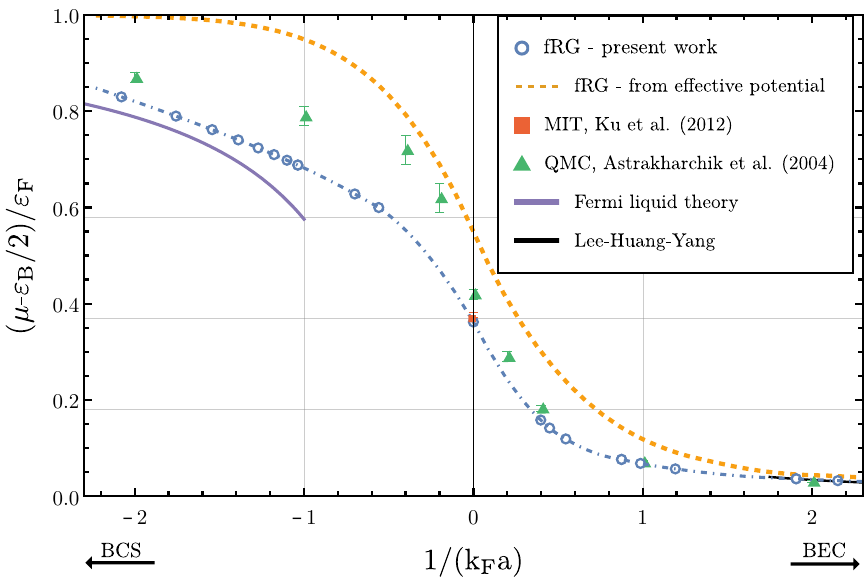}
		\caption{Equation of state from the present fRG approach (blue open circles and dash-dotted blue line) in comparison to that directly obtained from the effective potential in the present approximation (dashed orange line). Further theoretical results: Fermi-liquid theory (FLT) in the BCS-regime (purple solid line), Lee-Huang-Yang results on the BEC-side (black solid line), quantum Monte Carlo calculations (green triangles) by \cite{Astrakharchik_2004}. Experimental data: (red square) by the Zwierlein group at MIT \cite{Ku_2012}.\hspace*{\fill}} 
		\label{fig:EoS}
	\end{subfigure}
	\hspace*{.51cm}
	\begin{subfigure}{.48\linewidth}
		\centering
		\includegraphics[width=.95\textwidth]{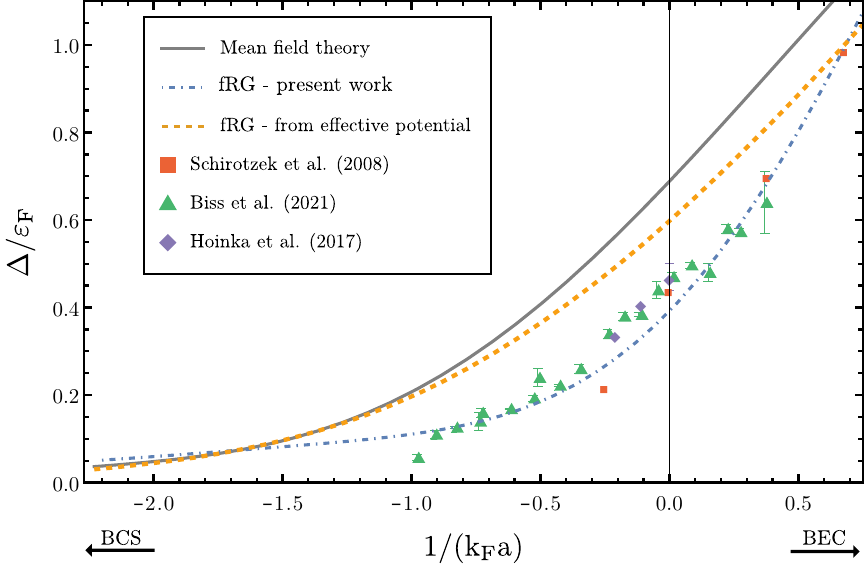}
		\caption{Gap from the present fRG approach (blue open circles and dash-dotted line), compared to the one obtained from mean-field calculations (solid grey line) and to that directly obtained from the effective potential in the present approximation (dashed orange line). In addition, we display experimental data from the Ketterle group at MIT \cite{Schirotzek_2008}, measured in a spin-imbalanced Fermi gas, shown as red squares, from the Moritz group in Hamburg \cite{biss_2021} in green triangles and by Hoinka et al. \cite{hoinka_2017} in purple diamonds.\vspace{.37cm}\hspace*{\fill}}
		\label{fig:Gap}
	\end{subfigure}
	\caption{EoS and gap at $T=0$ in comparison to experimental data and theoretical results in the literature.   }
	\label{fig:EoS+Gap}
\end{figure*}
%

%%%%%%%%%%%%%%%%%%%%%%%%%%%%%%%%%%%
\subsection{Equation of state}
\label{sec:EoS}

The result for the equation of state at zero temperature over the three-dimensional BCS-BEC crossover is shown in \Cref{fig:EoS}. For the Bertsch parameter $\xi=\mu/\epsilon_F$ at unitarity with $1/(k_f a)\to 0$, we obtain $\xi_{\text{fRG}}\simeq 0.362$ in good quantitative agreement with the experimental value of $\xi_{\text{exp}}=0.376(5)$ from \cite{Ku_2012}. In \Cref{fig:EoS} we also display the fRG result obtained directly from the effective potential within the present approximation. This demonstrates very clearly, that the discrepancy for the Bertsch parameter is dominated by the deficiencies of the previous extraction of density-related observables directly from the effective potential or from the integrated flows without resolving the fine-tuning problem for the initial conditions as discussed in \Cref{sec:density}. 

The results of the current work are compared with quantum Monte Carlo calculations by \cite{Astrakharchik_2004} and the experimental data by the Zwierlein group at MIT \cite{Ku_2012}. We find good quantitative agreement at unitarity and on the BEC-side with theory calculations and experimental results. On the BCS-side of the crossover, our result seems to yield smaller values for the equation of state than comparable calculations, although it agrees well with the Fermi liquid theory result for sufficiently large negative scattering length, $k_f a \to -\infty$. Furthermore, there is good agreement with various other calculations from quantum Monte Carlo simulations \cite{chang_2004,pilati_2008}, as well as T-matrix \cite{pieri_2005} and Nozières, Schmitt-Rink \cite{hu_2006} studies and measurements \cite{navon_2010}.

%%%%%%%%%%%%%%%%%%%%%%%%%%%%%%%%
\subsection{Gap}
\label{sec:Gap}

The results for the gap over the Fermi energy, $\Delta/\epsilon_F$, at zero temperature over the three-dimensional BCS-BEC crossover is shown in  \Cref{fig:Gap}.  At unitarity with $1/(k_f a)\to 0$, we obtain $(\Delta/\epsilon_F)_{\text{fRG}}\simeq 0.39$ in reasonable agreement with the experimental value of $(\Delta/\epsilon_F)_{\text{exp}}=0.44$ from \cite{Schirotzek_2008}. We compare the gap calculated in the present work to the fRG result obtained from the effective potential calculation, the gap in mean field theory, experimental data by the Ketterle group at MIT \cite{Schirotzek_2008} (obtained in a spin-imbalanced Fermi gas), by the Moritz group in Hamburg \cite{biss_2021} and by \cite{hoinka_2017}. We find good quantitative agreement with our fRG results over the whole crossover regime. In addition, good agreement is found with beyond mean field calculations in \cite{pisani_2018}, functional integral approaches \cite{diener_2008,gezerlis_2008,hausmann_2009,chen_2016,tajima_2017}, quantum Monte Carlo simulations \cite{bulgac_2008} and measurements in \cite{behrle_2018}.

%%%%%%%%%%%%%%%%%%%%%%%%%%%%%%%%%%
\section{Conclusion}

In the present work we have complemented the fRG approach for ultracold atomic systems as well as other non-relativistic theories with a computational framework for the evaluation of density-related observables. This allowed us to compute, for the first time in the fRG approach, the Bertsch parameter and the gap of ultracold fermionic gases with semi-quantitative precision already within a simple approximation. Importantly, this analysis also corroborates the semi-quantitative reliability of the fRG approach for the dynamics of ultracold atomic systems in simple approximations. 

We have shown that the discrepancies observed so far were solely related to the ultraviolet fine-tuning problem for \textit{derived} observables such as the density $n$ as well as density fluctuations. Derived observables do not feed back into the flow of the effective action, that is in the flow of the 1PI correlation functions. The density and its low order $\mu$-derivatives show a $\mu$-dependence which scales with positive powers of the cutoff scale $k$. This problem can be accommodated as $\partial_\mu^i n$ as well as its flow go to zero in the ultraviolet for $i\geq 2$ and appropriately chosen regulators. The first decaying hyper fluctuation is the skewness. Then, the lower order fluctuations and the density are obtained by integrations over the chemical potential.  

In turn, for given $\mu$, the coupling parameters are not subject to the above fine-tuning problem. In combination with the semi-quantitative results for the density fluctuation observables this provides semi-quantitative reliability for the 1PI correlation functions in ultracold gases, scaled with appropriate powers of the density. We emphasise that the current results also apply to other non-relativistic and relativistic systems. The reliable computation of $n(\mu)$ opens the door for precision predictions of many observables through fRG computations within advanced truncations.

%%%%%%%%%%%%%%%%%%%%%%%%%%%%%%%%%%
\section{Acknowledgements}

We thank Igor B\"ottcher, Stefan Fl\"orchinger and Selim Jochim for discussions. This work is funded by the Deutsche Forschungsgemeinschaft (DFG, German Research Foundation) under Germany’s Excellence Strategy EXC 2181/1 - 390900948 (the Heidelberg STRUCTURES Excellence Cluster) and the Collaborative Research Centre SFB 1225 - 273811115 (ISOQUANT).

%%%%%%%%%%%%%%%%%%%%%%%%%%%%%%%%%%%%%%%%%%%%%%%%%%%%%%%%

\appendix

%%%%%%%%%%%%%%%%%%%%%%%%%%%%%%%%%%
\section{Initial conditions}\label{app:initial}

The microscopic parameters in the effective action \labelcref{eq:Gk} are the Yukawa coupling $h$ and the bosonic parameters $m_\phi^2, \lambda_\phi$ in the effective potential. The ratio $S_\phi$ of temporal and spatial wave function is set to unity in the ultraviolet. The parameters are summarised in \Cref{tab:MicroParameters}. In this context we pay particular attention to the choice of the infrared regulators. Some regulators, as a sharp cutoff around the Fermi surface, are rather useful for understanding the rôle of fluctuations with momenta around the Fermi momentum or below as well as incorporating these fluctuations in an efficient way. However, in the ultraviolet they typically induce a qualitatively larger growth in density-related correlations such as $U_0$ which should be avoided.

\begin{table}[h]
	\begin{center}
		\begin{tabular}{ |c|c|c|c|c|c|}
			\hline & & &&& \\[-1ex]
			$g_{i}$ & $m_{\phi}^2$ & $\lambda_{\phi}$ & $\rho_{0}$& $S_{\phi}$& $h^2$  \\[1ex] 
			\hline  & & &&& \\[-2ex]
			& $\nu_{\Lambda} - 2\,\mu$ & $\tilde{\lambda}_{\phi, *}/\Lambda$  &0 & 1& $6\,\pi^2\,\Lambda$ \\[1ex] 
			\hline
		\end{tabular}
		\caption{Summary of the initial conditions for the couplings $g_i$. The subscript ${}_*$ in $\tilde{\lambda}_{\phi, *}$ indicates the fixed point for the dimensionless coupling $\lambda_{\phi}$ and $\nu_{\Lambda}$ the detuning at $k=\Lambda$.\hspace*{\fill}}
		\label{tab:MicroParameters}
	\end{center}
\end{table}
%

%%%%%%%%%%%%%%%%%%%%%%%%%%%%%%%%%%%%
\section{Flows and fine-tuning}
\label{app:TechnicalDetails}

In this Appendix we provide the explicit expressions for the flow equation of the effective potential and the wave functions. For more details see~\cite{Faigle-Cedzich:2019uvr} and the reviews ~\cite{Kopietz:2010zz, Metzner:2011cw, Scherer:2010sv, Boettcher:2012cm, Dupuis:2020fhh}.

We also discuss the UV fine-tuning problem triggered by $\mu$-dependent terms with relevant UV scaling with positive powers of $k$.

%%%%%%%%%%%%%%%%%%%%%%%%%%%%%%
\subsection{Flows with Fermi surface regulators}
\label{app:FermisurfaceRegulator}
	
The infrared cutoff is introduced via a modification of the dispersion in the classical action \labelcref{eq:action_HS}, 
\begin{align}
	\int_Q \,\Bigl[ \psi(-Q)\,R_{\psi}(Q)\,\psi(Q)+\phi(-Q)\,R_{\phi}(Q)\,\phi(Q) \Bigr]\,, 
\end{align}
where the regulators $R_{\psi/\phi}(Q)$ have the limits 
\begin{align}
	\lim_{q^2/k^2\rightarrow 0}\,R_{\psi/\phi}(Q)=k^2\,,\quad \lim_{q^2/k^2\rightarrow \infty}\,R_{\psi/\phi}(Q)=0\,. 
\end{align}
In our explicit computations we use a spatial flat regulator \cite{2001IJMPA..16.2081L,Litim:2001up,PhysRevB.77.064504} for sufficiently low cutoff scales $k<k_0$. The scale $k_0$ will be discussed later in \Cref{app:CombinedRegulator}, and is chosen for optimising both the inclusion of the infrared physics as well as the decoupling of the ultraviolet scaling part. In the infrared we choose 
\begin{align}\nonumber
	R^\textrm{\tiny{IR}}_{\psi}(q^2)= & k^2\,\left(\sgn\left(z\right) -z\right)\,\theta\left(1-|z|\right)\,,\\[1ex]
		R_{\phi}(q^2)= &\left(k^2-q^2/2\right)\,\theta\left(k^2-q^2/2\right)\,,
	\label{eq:flat_reg}
\end{align}
where $\theta(x)$ is the Heaviside step function, $\text{sgn}(x)$ is the sign function and 
\begin{align}
z=(q^2-\mu)/k^2\,. 
\label{eq:z}
\end{align}
With the regulators \labelcref{eq:flat_reg}, the flow equation of the effective action in $d$ space dimensions takes  the form 
\begin{align}
\dot{U}(ρ) &= - \frac{16\,v_d}{d}\,k^{d+2}\,\ell_F^{(1,1)}+\frac{8\,v_d\,2^{d/2}}{d} \,k^{d+2}\,\ell_B^{(1,1)}\,, 
\label{eq:eff_pot_flow_opt}	
\end{align}
with the fermionic and bosonic threshold functions $\ell_F^{(1,1)}$ and $\ell_B^{(1,1)}$ provided in \labelcref{eq:threshold1,eq:ellBnm} in \Cref{app:ThresholdFunctions}. The factor $v_d$ comprises the $d$-dimensional angular integral and is given by 
\begin{align}
v_d  = \frac{1}{2^{d+1}\,\pi^{d/2}\,\Gamma(d/2)}\,.
\end{align}

\begin{widetext}
The temporal and spatial anomalous dimensions of the dimer are given by 
\begin{align}\nonumber 
		\dot{S}_ϕ =&\, - \frac{16\,h^2\,v_d}{d}\,k^{d-4}\left(\ell_F^{(0,2)} -2\,w_3\,\ell_F^{(0,3)}\right)-\frac{32\,S_ϕ}{d}\,ρ\,U^{(2)}\,v_d\,2^{d/2}\,k^{d-4}\\[1ex] \nonumber 
		& \times \Biggl[
		\left(U^{(2)} +ρ \,U^{(3)}\right) \ell_B^{(0,2)}+2\left(ρ\, U^{(2)}\right)^2 \left(U^{(2)}+ρ\, U^{(3)}\right) k^{-4} \ell_B^{(0,3)} -2ρ \,U^{(2)}\left(2\,U^{(2)} +ρ\,U^{(3)}\right)\,k^{-2}\,\ell_B^{(1,3)}
		\Biggr]\,,\\[2ex]
		η_ϕ &=  \frac{16\,h^2\,v_d}{d}\,k^{d-4}\,\ell_{F,2}^{(0,2)}+8\,ρ\,\left(U^{(2)}\right)^2\frac{v_d\,2^{d/2}}{d}\,k^{d-4}\,\ell_{B,2}^{(0,2)}\,.
\label{eq:runningSphiAphi}
\end{align}
The fermionic and bosonic threshold functions $\ell_F$ and $\ell_B$ are provided in \labelcref{eq:threshold1,eq:threshold2,eq:ellBnm,eq:ellB2nm} in \Cref{app:ThresholdFunctions}. They involve the anomalous dimensions $\eta_\phi$ and the ratio $S_\phi$ of the spatial and temporal wave function. 
\end{widetext}

%%%%%%%%%%%%%%%%%%%%%%%%%%%%%%
\subsection{UV Flows with spatial momentum regulators}
\label{app:SpatialRegulator}

In the ultraviolet the Fermi surface does not matter for the dynamics and for cutoff scales $k>k_0$ we choose a fermionic spatial momentum cutoff, 
\begin{align}
	R_\psi^{\text{\tiny{UV}}}(q^2)=k^2\,\left(1-z\right)\,\theta\left(1-z\right)\,, 
	\label{eq:Ruv}
\end{align}
while the bosonic regulator in \labelcref{eq:flat_reg} is used throughout. This only changes the fermionic parts in the flows \labelcref{eq:eff_pot_flow_opt,eq:runningSphiAphi}, 
\begin{align}
\dot{U}^{(F)}(ρ) &= - \frac{16\,v_d}{d}\,k^{d+2}\,\ell_{F,\text{UV}}^{(1)}\,,
\label{eq:FermFlowsRspatial}
\end{align}
where the superscript ${}^{(F)}$ indicates the contribution of the fermionic loop. The modified threshold function is
given by 
\begin{align}\nonumber
\ell_{F,\text{UV}}^{(m)} = \left(1+ \tilde{\mu}\right)^{d/2}\,\theta\left(1+\tilde{\mu}\right)\,\mathcal{F}_m\left(\sqrt{1+w_3}\right)\,,
\end{align}
with $\mathcal{F}_m$ defined in \Cref{app:ThresholdFunctions}.

%%%%%%%%%%%%%%%%%%%%%%%%%%%%%%%%%%%%
\subsection{Ultraviolet scaling of Fermi surface flows}
\label{app:UV-Scaling}

In contrast to the spatial momentum regulator, the Fermi surface regulator in \labelcref{eq:flat_reg} induces an additional $\mu$-dependent enhancement factor in the ultraviolet regime for large $k$. Hence, in this regime it should not be used. We demonstrate this by the rôle played by $S_\phi$, which should tend towards unity in the UV, or, more generally, scales as $k^0$. The fermionic part $\dot{S}_\phi^{(F)}$ of its flow is given by the first line of \labelcref{eq:runningSphiAphi}. For $d=3$ spatial dimensions and vanishing temperature it is proportional to 
\begin{align}
	\dot{S}_\phi^{(F)} \propto  \frac{h^2}{k}\,\left[\theta({\tilde\mu}+1)\,({\tilde\mu}+1)^{3/2}-2\,\theta({\tilde\mu})\,{\tilde\mu}^{3/2}\right]\,,
\label{eq:sphiEq}\end{align}
with the dimensionless chemical potential 
\begin{align}
\tilde \mu=\frac{\mu}{k^2}\,. 
\end{align}
Here, we have dropped $k$-independent prefactors, and have used $\phi=0$ in the symmetric phase and  $k^2\gg μ$. Importantly, the term $2\,\theta({\tilde\mu})\,{\tilde\mu}^{3/2}$ originates in the sign-term in Fermi surface regulator \labelcref{eq:flat_reg}. This term is absent for the momentum regulator in \labelcref{eq:Ruv}. 

The flow of the skewness, $\dot n^{(2)}$, is obtained by taking three $μ$-derivatives of the flow of the effective potential, that includes a part that is proportional to $\dot{S}_\phi$. Hence, $\dot n^{(2)}$ includes terms with 
\begin{align}
k^5	∂_μ^3\dot{S}_\phi^{(F)} \propto \frac{h^2}{k} \frac{1}{k} \left[\frac{\theta({\tilde\mu}+1)}{({\tilde\mu}+1)^{3/2}}-2\,\frac{\theta({\tilde\mu})}{{\tilde\mu}^{3/2}}\right]\,,
\end{align}
where the prefactor $k^5$ is the canonical dimensional scale factor $k^5$ due to the momentum trace in the flow of the effective action, $\dot\Gamma_k$, or that of the effective potential, $\dot U$. The first term in the square bracket scales with $k^0$, while the second term scales with $k^3$. This term is only present for Fermi surface regulators, and in comparison we find the asymptotic UV-behaviour 
\begin{align}
k^5	∂_μ³ {\dot S}_\phi ∝
	\left\{\begin{array}{rl}
	h^2\,k &  \quad \text{Fermi surface regulators}\\[1ex]
	\frac{h^2}{ k^2} & \quad \text{otherwise}
\end{array}\right.\,.
\label{eq:SPhiBehaviourUV}
\end{align}
We remark for completeness, that the Yukawa or Feshbach coupling $h_k$ is attracted to a (partial) fixed point in the UV and has the scaling behaviour $h_k \sim \sqrt{k}$ in the ultraviolet which is ignored in the present computation. If this improvement of the approximation is taken into account, the UV-scaling in \labelcref{eq:SPhiBehaviourUV} is multiplied by $k$, and the fine-tuning problem is amplified. 

In any case, we infer from \labelcref{eq:SPhiBehaviourUV}, that the flow of the skewness $n^{(2)}$ contains a term 
\begin{align}
	S_\phi^{(3)}\,\frac{ \partial \dot{Γ}_k}{ \partial S_\phi } ∝
	\left\{\begin{array}{rl}
		h^2\,k &  \quad \text{Fermi surface regulators}\\[1ex]
		\frac{h^2}{ k^2} & \quad \text{otherwise}
	\end{array}\right.\,.
\end{align}
In summary, with a Fermi surface regulator in the UV, we encounter a fine-tuning problem in the UV, which is absent otherwise. Note also that the fine-tuning problem for Fermi surface regulators is even amplified in improved approximations, while it is still absent for other regulators. This scaling analysis also holds true at finite temperatures, as the flows reduce to the vacuum flows for $T/k^2\rightarrow 0$. 

\begin{figure}[t]
	\centering
	\includegraphics[width=\columnwidth,valign=t]{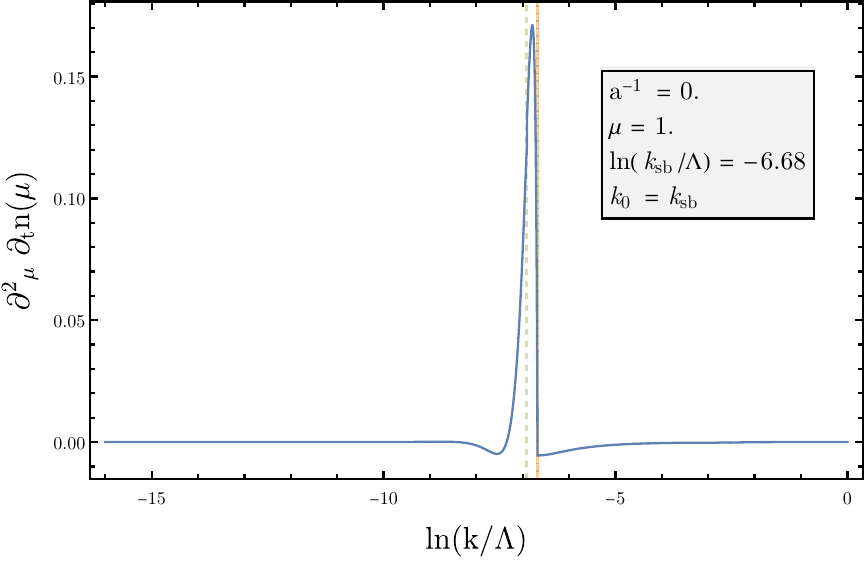}
	\makeatletter\long\def\@ifdim#1#2#3{#2}\makeatother
	\caption{The flow of the second $\mu$-derivative of the density, here exemplary shown for a diverging scattering length of $a^{−1} = 0$ and $\mu = 1$, is peaked around the switching scale $k_0 = k_{\text{sb}}$. The solid orange line depicts the symmetry breaking scale $\ln(k_{\text{sb}}/\Lambda)$, while the green dashed line shows the chemical potential
		$\ln(\mu/\Lambda)$ and the grey dotted one the switching scale $\ln(k_0/\Lambda)$ which coincides with the symmetry breaking scale here.}
	\label{fig:2ndDerFlow}
\end{figure}
%

%%%%%%%%%%%%%%%%%%%%%%%%%%%%%%%%%
\subsection{Flows with UV-IR improved regulator} 
\label{app:CombinedRegulator}

In \Cref{app:UV-Scaling} we have shown that Fermi-surface regulators trigger UV fine-tuning problems that are absent for other regulators. However, the relevant infrared physics is best described in terms of momentum fluctuations about the Fermi surface. Accordingly, we introduce UV-IR improved regulators that lack the UV fine-tuning problem, while still providing a Fermi-surface regularisation in the infrared. 

This is done by a fermionic regulator that interpolates between a Fermi surface cutoff in the infrared such as $R_{\psi,k}(q^2)$ in \labelcref{eq:flat_reg} and a spatial momentum cutoff with $\sgn\left(z\right)\to 1$ in the ultraviolet. A simple choice is given by 

\begin{align}
  R_\psi=R_\psi^{\text{\tiny{UV}}}\,\theta(k-k_0)+R_\psi^{\text{\tiny{IR}}}\,\theta(k_0-k)\,,
\label{eq:CombinedRegulator} 
\end{align}
with the UV regulator \labelcref{eq:Ruv} and the Fermi surface regulator in \labelcref{eq:flat_reg}. At the scale $k_0$ we switch between ultraviolet and infrared regulator. While we choose this scale close to the Fermi surface, the results for $k\to 0$ should not depend on this choice as long as $k_0$ is larger than the Fermi surface and not asymptotically large. The advantage of employing a sharp switching between the infrared and ultraviolet regulator is the possibility to analytically perform all integrations and Matsubara sums. For a smooth switching behaviour one would have to deal with both the influence of the switching function itself, as well as the contributions of the infrared and ultraviolet part in a certain switching region. 

The scale derivative of a fermionic regulator $R_\psi$ with \labelcref{eq:CombinedRegulator}  reads 
\begin{align}\nonumber 
  \dot{R}_\psi= &\,\dot R_\psi^{\text{\tiny{UV}}}\,\theta(k-k_0)+\dot R_\psi^{\text{\tiny{IR}}}\,\theta(k_0-k)\\[1ex]
  & +k\,\delta(k-k_0)\,\left( R_\psi^{\text{\tiny{UV}}}-R_\psi^{\text{\tiny{IR}}}\right) \,,
\label{eq:SplitRegDer}
\end{align}
The combined regulator \labelcref{eq:CombinedRegulator} appropriately regularises around the Fermi surface in the infrared. The respective propagator reads 
\begin{align}
  G_{\psi}^Q=G_{\psi}^{\text{\tiny{UV}}}\, \theta(k-k_0)+G_{\psi}^{\text{\tiny{IR}}}\,\theta(k_0-k)\,.
\label{eq:FermionPropagatorSplit}
\end{align}
Thus, the flow equations for the effective potential and for the couplings, obtained via appropriate projection descriptions split up into an UV- and IR-part, that only depend on the respective regulators, as well as a term which is proportional to $( R_\psi^{\text{\tiny{UV}}}-R_\psi^{\text{\tiny{IR}}})\,\delta(k-k_0)$. The last term has only (analytic) contributions at the switching scale $k_0$. In practice, we choose the switching scale as small as possible, i.e. the symmetry breaking scale $k_0=k_{\text{sb}}$ on the BCS-side of the crossover and at unitarity. On the BEC-side, we choose $k_0=2/3\,k_{\text{sb}}$, since the the symmetry breaking occurs already at much larger scales during the flow. 

For illustration we show the flow of the skewness $n^{(2)}$ in  \Cref{fig:2ndDerFlow} at the Feshbach resonance with a diverging scattering length $a^{−1} = 0$. We have checked that our results do not depend on the switching scale, and are also insensitive to changes to soft switching. Note, that this is not surprising in the present approximation, where the fermionic loop can be written as a total $t$-derivative. Accordingly, one can show analytically, that the fermionic loop is regulator-independent. Only the bosonic feedback may lead to an approximation dependence. 

The results shown in the present work are obtained for  $k_0=k_{\text{sb}}$. Here, $k_0$ is the scale, where we switch from a spatial momentum cutoff in the ultraviolet to a Fermi surface cutoff in the infrared, see \Cref{app:CombinedRegulator}. We emphasise, that the results are independent of $k_0$ as long as $k_0$ is not deep in the UV and IR asymptotic regime, and this independence has also been checked.

%%%%%%%%%%%%%%%%%%%%%%%%%
\subsection{Threshold functions} 
\label{app:ThresholdFunctions}

Here we used the definitions for fermionic contributions 
\begin{align}
	\ell_F^{(n,m)}\left(\tilde{μ},\tilde{T},w_3\right) =
	\begin{cases}
		\ell_2(\tilde{μ})\,\mathcal{F}_{\!\!m}\!\!\left(\sqrt{1+w_3}\right)\quad n\,\,\,\text{even}\\[1ex]
		\ell_1(\tilde{μ})\,\mathcal{F}_{\!\!m}\!\!\left(\sqrt{1+w_3}\right)\quad n\,\,\,\text{odd}
	\end{cases}\,,
\label{eq:threshold1}
\end{align}
and
\begin{align}
	\ell_{F,2}^{(n,m)}\left(\tilde{μ},\tilde{T},w_3\right) =
	\begin{cases}
		\ell_3(\tilde{μ})\,\mathcal{F}_{\!\!m}\!\!\left(\sqrt{1+w_3}\right)\quad n\,\,\,\text{even}\\[1ex]
		\ell_1(\tilde{μ})\,\mathcal{F}_{\!\!m}\!\!\left(\sqrt{1+w_3}\right)\quad n\,\,\,\text{odd}
	\end{cases}\,,
\label{eq:threshold2}
\end{align}
where we made use of $w_1=V'/k^2$ and $w_2=ρ\,V^{(2)}/k^2$, as well as $w_3=h^2\,ρ/k^4$. For bosonic diagrams we defined
\begin{align}\nonumber 
	\ell_B^{(n,m)}\left(\tilde{T},w_1,w_2\right) =& \frac{1}{S_ϕ^{2m}}\,\left(1-\frac{η_ϕ}{d+2}\right)\,\left(1+w_1+w_2\right)^n\\[1ex]
&\hspace{-2cm}\times 	\mathcal{B}_{m}\left(\sqrt{(1+w_1)(1+w_1+2\,w_2)}/S_ϕ\right)\,,
\label{eq:ellBnm}
\end{align}
and
\begin{align}
	\ell_{B,2}^{(0,m)}\left(\tilde{T},w_1,w_2\right) = \left.\ell_B^{(0,m)}\right\rvert_{η_ϕ=0}\,. 
	\label{eq:ellB2nm}
\end{align}
$\mathcal{F}_{m}(z)$ and $\mathcal{B}_{m}(z)$ are the fermionic and bosonic Matsubara sums of order $m$ given in \labelcref{eq:Mat_F,eq:Mat_B}, respectively. The functions $\ell_i$ are given by 
\begin{align}\nonumber 
		\ell_1(x) = &\, θ(x+1)\,(x+1)^{d/2} - θ(x-1)\,(x-1)^{d/2}\,,\\[1ex]\nonumber 
		\ell_3(x) = &\,θ(x+1)\,(x+1)^{d/2} + θ(x-1)\,(x-1)^{d/2}\,,\\[1ex]
		\ell_2(x) =&\, \ell_3(x)-2\,θ(x)\,x^{d/2}\,.
\label{eq:ell_i}
\end{align}
%

%%%%%%%%%%%%%%%%%%%%%%%%%%%%%%%%%%%%
\subsubsection{Fermionic Matsubara sums}

For fermionic Matsubara sums we find
\begin{align}\nonumber 
		\mathcal{F}_1(z) &= T\,\sum_{ε_n}\,\frac{1}{ε_n^2+z^2}= \frac{1}{z}\,\left[\frac{1}{2}-n_F(z)\right]\,,\\[1ex]
			\mathcal{F}_n(z) &= -\frac{\partial}{\partial z^2}\,\mathcal{F}_{n-1}(z) \,,
	\label{eq:Mat_F}
\end{align}
with the Fermi-Dirac distribution 
\begin{align}
	\label{eq:FermiDirac}
	n_F(z) =
	\frac{1}{e^{\frac{z k^2}{T}} + 1}\,. 
\end{align}
For $T\to 0$ the Fermi-Dirac distribution tends towards $\theta(-z)$.

%%%%%%%%%%%%%%%%%%%%%%%%%%%%%%%%%
\subsubsection{Bosonic Matsubara sums}

For the bosonic Matsubara sum we find 
\begin{align}\nonumber 
		\mathcal{B}_1(z) &= T\,\sum_{ω_n}\,\frac{1}{ω_n^2+z^2}= \frac{1}{z}\,\left[\frac{1}{2}+n_B(z)\right]\,,\\[1ex]
	\mathcal{B}_n(z) &= -\frac{\partial}{\partial z^2}\,\mathcal{B}_{n-1}(z) \,,
\label{eq:Mat_B}
\end{align}
with the Bose-Einstein distribution
\begin{align}
		\label{eq:BoseEinstein}
		n_B(z) =
		\frac{1}{e^{\frac{z k^2}{T}} - 1}\,. 
\end{align}
For $T\to 0$ the Bose-Einstein distribution tends towards $-\theta(-z)$.

%%%%%%%%%%%%%%%%%%%%%%%%%%%%%%%%%%
\section{Density from $\mu$-integration of hyper fluctuations}
\label{app:density}

The density is computed via the integration of the skewness $\partial_\mu^2 n$, see	\labelcref{eq:Intn}. This requires the computation of $\partial_{\mu}^2n_k(\mu)$ from its flow. We start from the flow of the density, 
\begin{align}
	\dot n_k=   \frac{d\dot \Gamma_k}{d\mu } = \frac{\partial \dot \Gamma_k}{\partial\mu } 
	+\frac{d  g_i }{d \mu } \frac{\partial \dot  \Gamma_k}{ \partial g_i }\,, 
	\label{eq:flown-rewrite}
\end{align}
where the partial $\mu$-derivative is at fixed $\boldsymbol{g}$. We also have introduced the notation $\dot f=\partial_t f$ for the sake of simplicity. Both terms in \labelcref{eq:flown-rewrite} follow analytically from the master equation, \labelcref{eq:flowG}, and each partial $\mu$-derivatives and $d g_i/d\mu\, \partial_{g_i}$-derivative lowers the effective $k$-dimension by two. The coefficients $d g_i/d\mu $ follow from their flow
\begin{align}
	\dot  g_i^{ (1) } = \frac{d \dot g_i}{d\mu}=\frac{\partial \dot g_i}{\partial \mu}  +
	g^{ (1) }_j\frac{\partial\dot g_i}{ \partial {g_j}}\,,
	\label{eq:flowmul}
\end{align}
where we have introduced the notation 
\begin{align}
	g^{(n)}_i=\frac{d^n g_i}{d\mu^n}\,,  \qquad g_i^{(0)}=g_i\,.
	\label{eq:mug_App}
\end{align}
\Cref{eq:flowmul} is a coupled differential equation for $\boldsymbol{g}^{(1)}$,
\begin{subequations}
	\label{eq:Flowg1}
	\begin{align}
		\dot{\boldsymbol{g}}^{ (1) }=\boldsymbol{A}_1 + B\, \boldsymbol{g}^{ (1) }\,,
		\label{eq:flowmuform}
	\end{align}
	with 
	\begin{align}
		\boldsymbol{A}_{1}(\mu;\boldsymbol{g})= \frac{\partial \dot g_i}{\partial \mu}   \,,\qquad B_{ij}(\mu;\boldsymbol{g})= \frac{\partial \dot g_i}{\partial {g_j}}\,.
		\label{eq:coeffs1}
	\end{align}
\end{subequations}
The coefficients $A_{1,i}$ and $B_{ij}$ can be read-off from the flow of the couplings $\boldsymbol{g}$, that are directly derived from \labelcref{eq:flowG}. The flows $\dot{\boldsymbol{g}}$ only depend on $\boldsymbol{g}$ and not on $\boldsymbol{g}^{(n>0)}$. Hence \labelcref{eq:flowmuform} is a \textit{derived} flow: it does not feed back into the flow of the effective action. Naturally, \labelcref{eq:flowmuform} this can be iteratively extended to the higher $\mu$-derivatives of $\boldsymbol{g}$. For $g_i^{(2)}$ it reads
\begin{align}\nonumber 	
	\dot  g_i^{ (2) } &=\frac{d}{d\mu}\left(A_{1,i}   +
	B_{ij}\, g^{ (1) }_j\right)\\[2ex]\nonumber 
	&= \partial_\mu A_{1,i} + g^{ (1) }_j\frac{\partial A_{1,i}}{\partial {g_j}}\\[1ex]
	& + g^{ (1) }_j \left( \partial_\mu+ g_m^{ (1) }\frac{\partial }{\partial {g_m}} \right) \,  B_{ij}+B_{ij}\, g^{ (2) }_j\,, 
	\label{eq:flowmul1}
\end{align}
of in terms of the full system of linear differential equations for $\boldsymbol{g}$, 
\begin{subequations}
	\label{eq:Flowg2}
	\begin{align}
		\dot{ \boldsymbol{g}}^{ (2) }=\boldsymbol{ A}_2 + B\,\boldsymbol{  g}^{ (2) }\,,
		\label{eq:flowmu2}
	\end{align}
	with the matrix $B$ in \labelcref{eq:coeffs1} and the coefficients $A_{2,i}$ of the vector $\boldsymbol{ A}_2$,  
	\begin{align}\nonumber 
		A_{2,i}(\mu;\boldsymbol{g}, \boldsymbol{g}^{(1)})=&\, \left( \partial_\mu+ g_m^{ (1) }\frac{\partial }{\partial {g_m}} \right) A_{1,i}\\[1ex] 
		&\hspace{-0cm}+ g^{ (1) }_j \left( \partial_\mu+ g_m^{ (1) }\frac{\partial }{\partial {g_m}} \right) B_{ij} \,,
		\label{eq:coeffs2form}
	\end{align}
	with $B$ given in \labelcref{eq:coeffs1} and 
	\begin{align}
		\left( \partial_\mu+ g_m^{ (1) }\frac{\partial }{\partial {g_m}} \right) B_{1,ij}=
		\partial_\mu\frac{\partial \dot g_i }{\partial {g_j}} +
		g^{ (1) }_m \frac{  \partial^2 \dot g_i }{ \partial {g_m}\partial {g_j}}\,.
		\label{eq:Bder}
	\end{align}
\end{subequations}
\Cref{eq:Flowg1,eq:Flowg2} are readily iteratively extended to the flow of a general $\boldsymbol{g}^{(n)}$: it satisfies the differential equation 
\begin{subequations}
	\label{eq:Flowgn}
\begin{align}
		\dot{ \boldsymbol{g}}^{ (n) }=\boldsymbol{ A}_n + B\,\boldsymbol{  g}^{ (2) }\,,
		\label{eq:flowmun}
\end{align}
with the matrix $B$ in \labelcref{eq:coeffs1} and  the vector $\boldsymbol{A}_n$. The vector $\boldsymbol{A}_n$ depends on $\mu$ and $g^{(m<n)}$. It can be determined iteratively from $\boldsymbol{A}_{n-1}$ and $B$, 
	\begin{align}\nonumber 
		A_{n,i}(\mu;\boldsymbol{g}, ..., \boldsymbol{g}^{(n-1)})=& 
		\left( \partial_\mu+
		\sum_{j=1}^{n-1} g_{m}^{(j)}
		\frac{\partial}{\partial_{g^{(j-1)}_{m}}}
		\right)\, A_{n-1,i}
		\\[2ex]
		& \hspace{-.8cm}+
		g^{ (n-1) }_j\left( \partial_\mu+ g_m^{ (1) }\frac{\partial }{\partial {g_m}} \right) \,  B_{ij}\,.
		\label{eq:Ait}
	\end{align}
\end{subequations}
Note that there are various forms for the coefficients $A_n$ and $B_n$. The above forms have the advantage that all derivatives w.r.t. $\mu$ and $g^{(n)}_i$ can be performed analytically, and displays the iterative structure, that entails that the flows of the $\boldsymbol{g}^{(n>0)}$ are derived flows and do not feed back into the flow of the $\boldsymbol{g}$, which determine the effective action $\Gamma_k$.  

We now use the same iterative structure and dependences for the flow of the density and its hyper fluctuations, starting from the flow of the effective action and that of the density, \labelcref{eq:flown-rewrite}. We are led to the 
\begin{subequations}
	\label{eq:flowni_App}
	\begin{align}
		\frac{d \dot{n}^{(i)} }{d \mu}=
		\left( \partial_\mu+
		\sum_{j=1}^{i} g_{m}^{(j)}
		\frac{\partial}{\partial_{g^{(j-1)}_{m}}}
		\right)\, \dot n^{(i-1)}\,,
		\label{eq:flowmoments_App}
	\end{align}
	with
	\begin{align}
		\dot n^{(m)}=\frac{d^m \dot n}{d\mu^m}\,,  \qquad n^{(-1)}=\dot\Gamma_k[0,\phi_\textrm{\tiny{EoM}}]\,.
		\label{eq:mun_App}
	\end{align}
\end{subequations}
The highest order $g^{(i)}$ in \labelcref{eq:flowmoments_App} originates in the $\boldsymbol{g}$-dependence of 
$\dot \Gamma_k$, leading to $\dot n^{(0)}=\dot n(\mu; \boldsymbol{g},\boldsymbol{g}^{(1)})$. This entails by iteration, that 
\begin{align}
	\dot n^{(i)} = \dot n^{(i)}(\mu;\boldsymbol{g},...,\boldsymbol{g}^{(i+1)})\,. 
\end{align}
Collecting all the results, we can compute the density at $k=0$ by integrating $n^{(2)}$ twice over the chemical potential, 
\begin{align}
	n(\mu)=\int_0^\mu\,d\mu_1\left[\int_0^{\mu_1}\,d\mu_2\, n^{(2)}(\mu_2)\right]\,,
	\label{eq:densityIntegrated}
\end{align}
where we have used that $n(0)$ and $\partial_\mu\,n(0)$ are vanishing. The skewness $n^{(2)}(\mu)$ is computed from its flow, 
\begin{align}
	n^{(2)}(\mu)=\int_\Lambda^0\,\frac{dk}{k}\,\dot{n}^{(2)}_k(\mu;\boldsymbol{g},...,\boldsymbol{g}^{(3)})\,,
\end{align}
for an initial condition ${n}^{(2)}_{\Lambda}\approx 0$ for sufficiently large $\Lambda$.

\bibliographystyle{apsrev4-2}
\bibliography{Literature.bib}

\end{document}